\documentclass[twocolumn,showpacs,amssymb,aps]{revtex4}
\usepackage{amsmath}
\usepackage{graphicx}

\begin{document}

\title{Thermal Operator and Dispersion Relation in QED at Finite
  Temperature and Chemical Potential}

\author{Ashok Das$^{a,b}$ and J. Frenkel$^{c}$}
\affiliation{$^{a}$ Department of Physics and Astronomy,
University of Rochester,
Rochester, NY 14627-0171, USA}
\affiliation{$^{b}$ Saha Institute of Nuclear Physics, 1/AF
  Bidhannagar, Calcutta 700064, INDIA}
\affiliation{$^{c}$ Instituto de F\'{\i}sica, Universidade de S\~ao
Paulo, S\~ao Paulo, SP 05315-970, BRAZIL}

\bigskip


\begin{abstract}

Combining the thermal operator representation with the dispersion
relation in QED at finite temperature and chemical potential, we
determine the complete retarded photon self-energy only from its
absorptive part at zero temperature. As an application of this method,
we show that, even for the case of a nonzero chemical potential, the
temperature dependent part of the one loop retarded photon
self-energy vanishes in $(1+1)$ dimensional massless QED.

\end{abstract}

\maketitle

In a series of recent papers \cite{silvana,silvana1,das,das1}, we have
shown how the thermal operator representation
\cite{espinosa,silvana2,silvana3}, which relates a Feynman graph at finite
temperature to the corresponding one at zero temperature both in the
imaginary time formalism \cite{kapusta,lebellac} as well as in the
real time formalism of closed time path \cite{dasbook}, can be used
profitably to study various questions of interest at finite
temperature. For example, using thermal operator representation,  the
cutting rules at finite temperature and
chemical potential can be directly obtained \cite{silvana} and the miraculous
cancellations observed earlier \cite{bedaque,dasbook} can be easily
understood. The thermal operator representation also clarifies the
meaning of the forward scattering amplitude description for the
retarded amplitudes at finite temperature \cite{silvana1} by relating them to the
corresponding forward scattering description at zero temperature. The
method also allows us \cite{das} to use the Schwinger proper time method
\cite{schwinger} to derive the hard thermal loop effective actions
\cite{braaten,frenkel} in a simple manner. Furthermore, this approach
clarifies the origin of many of the distinguishing features of hard
thermal loop effective actions in gauge theories by tracing these
properties directly to the corresponding zero temperature theory
\cite{das1}.  

In this brief report, we present yet another example of
how the thermal operator representation can be combined with other
powerful tools in quantum field theory to obtain nontrivial results at
finite temperature and chemical potential. Specifically, we will show
that when combined with
dispersion relations, the thermal operator representation can lead
directly to the complete retarded self-energy at finite temperature
and chemical potential from a knowledge of only the absorptive part of
the retarded self energy at zero temperature. Although this can be
done for any theory, we will restrict ourselves to the retarded photon
self-energy in QED which is of much interest in the study of linear
response theory \cite{kapusta,lebellac}.

Dispersion relations have been studied extensively at zero temperature
\cite{bjorken}. For a retarded function $f(t) = \theta (t) f(t)$, the dispersion
relations arise from the fact that the function in the Fourier
transformed space can be written as
\begin{equation}
f (\omega, \vec{p}\,) = \frac{1}{2\pi i} \int d\omega'\ \frac{f
  (\omega', \vec{p}\,)}{\omega'- \omega - i\epsilon},
\end{equation}
which leads to the relations between the real and the imaginary parts
as 
\begin{eqnarray}
{\rm Re}\ f (\omega,\vec{p}\,) & = & \frac{1}{\pi} \int d\omega'\ \frac{{\rm
    Im}\ f(\omega',\vec{p}\,)}{\omega'- \omega},\nonumber\\ 
{\rm Im}\ f (\omega,\vec{p}\,) & = & - \frac{1}{\pi} \int d\omega'\ \frac{{\rm Re}\
  f(\omega',\vec{p}\,)}{\omega'- \omega}.\label{dispersion}
\end{eqnarray}
These relations, which are conventionally known as the dispersion
relations, can also be combined into one single relation
\begin{equation}
f (\omega,\vec{p}\,) = \frac{1}{\pi} \int d\omega'\ \frac{{\rm Im}\ f
  (\omega',\vec{p}\,)}{\omega'- \omega - i\epsilon},\label{absorptive}
\end{equation}
which determines the complete retarded amplitude at zero temperature
from a knowledge of only its absorptive part. Of course, relations
\eqref{dispersion} and \eqref{absorptive} are meaningful only if ${\rm
  Im} f(\omega,\vec{p}\,)$ vanishes for large values of $\omega$. If
it does not, one can have a subtracted relation (for simplicity of
notation, we will suppress the momentum arguments which should be
understood) 
\begin{eqnarray}
\left(f (\omega) - f (\omega_{0})\right) & = &
\frac{1}{\pi} \int d\omega'\left({\rm Im} (f (\omega') - f
  (\omega_{0})\right)\nonumber\\
 & & \times\left[\frac{1}{\omega'-\omega-i\epsilon} -
  \frac{1}{\omega'-\omega_{0}-i\epsilon}\right]\!\!,
\end{eqnarray}
where $\omega_{0}$ is an arbitrary subtraction point that is normally
chosen to be $\omega_{0}=0$ in the absence of a chemical potential.

We note, however, that for the purposes of a thermal operator
representation, only an unsubtracted relation such as in \eqref{absorptive}
will suffice. This is easily seen from the fact that the thermal
operator acts at the integrand level before the integration over
internal momenta are carried out
\cite{espinosa,silvana2,silvana3}. Since the absorptive part of the
self-energy involves a combination of delta functions with the
external energy  $\omega$ as
one of the arguments (it represents an on-shell process), for a fixed
value of the internal momentum, it vanishes for large values of
$\omega$ (the divergences arise only when the internal momenta are
integrated). The important thing to note is that the thermal operator,
which relates the finite temperature graphs to the zero temperature
ones, is real and, consequently, it maintains the real and the
imaginary nature of parts of an amplitude. Therefore, if $f (\omega) =
\varPi_{\rm R}^{(0, \mu)} (\omega)$
represents the retarded self-energy in a theory at zero temperature and nonzero
chemical potential $\mu$ at the integrand level (before
the internal momentum integrations are done), then by applying the
thermal operator, the dispersion relation at finite temperature and
nonzero chemical potential follows from \eqref{absorptive} to be (we
are suppressing the momentum arguments for simplicity)
\begin{equation}
\varPi_{\rm R}^{(T, \mu)} (\omega) = \frac{1}{\pi} \int d\omega'\ \frac{{\rm Im}\ 
  \varPi_{\rm R}^{(T,\mu)} (\omega')}{\omega'- \omega -
  i\epsilon},\label{absorptiveT}
\end{equation}
where we have identified
\begin{eqnarray}
\varPi_{\rm R}^{(T,\mu)} (\omega) & = & {\cal O}^{(T,\mu)} \varPi_{\rm
  R}^{(0,\mu)}
(\omega), \nonumber\\
{\rm Im}\ \varPi_{\rm R}^{(T,\mu)} (\omega) & = & {\cal
  O}^{(T,\mu)} {\rm Im}\ \varPi_{\rm R}^{(0,\mu)} (\omega),\label{tor}
\end{eqnarray}
with ${\cal O}^{(T,\mu)}$ denoting the appropriate thermal operator
for the amplitude \cite{silvana2,silvana3}. This generalizes the
dispersion relation 
\eqref{absorptive} at zero temperature to that at finite temperature
and chemical potential. Furthermore, through the use of the dispersion
relation and the thermal operator, this method
shows how the complete retarded self-energy
at finite temperature and chemical potential can be obtained from a
knowledge of only the absorptive part of the zero temperature retarded
self-energy.

Let us now demonstrate how this works in QED with a nonzero chemical
potential $\mu$ by calculating the retarded self-energy for the photon. The
Lagrangian density for the theory is given by
\begin{equation}
{\cal L} = - \frac{1}{4} F_{\mu\nu}F^{\mu\nu} + i \bar{\psi}
D\!\!\!\!\slash \psi - m \bar{\psi}\psi + \mu \bar{\psi}\gamma^{0}\psi,
\end{equation}
where $D_{\mu}$ denotes the covariant derivative and $F_{\mu\nu}$ is the
Abelian field strength tensor. In the closed time path formalism, the
propagator in the mixed space becomes a $2\times 2$ matrix and at zero
temperature has the form \cite{silvana2}
\begin{widetext}
\begin{eqnarray}
iS_{++}^{(0,\mu)} (t,\vec{p}\,)& = & \frac{e^{i\mu
    t}}{2E_{p}}\left(\theta (t) A(E_{p}) e^{-iE_{p}t} + \theta (-t) B(E_{p})
  e^{iE_{p}t}\right),\quad
iS_{+-}^{(0,\mu)} (t,\vec{p}\,) =  \frac{e^{i\mu t}}{2E_{p}}\ B(E_{p})
e^{iE_{p}t},\nonumber\\
iS_{-+}^{(0,\mu)} (t,\vec{p}\,) & = & \frac{e^{i\mu t}}{2E_{p}}\ A(E_{p})
e^{-iE_{p}t},\quad 
iS_{--}^{(0,\mu)} (t,\vec{p}\,)  =  \frac{e^{i\mu
    t}}{2E_{p}}\left(\theta (t) B(E_{p}) e^{iE_{p}t} + \theta (-t) A(E_{p})
  e^{-iE_{p}t}\right),\label{prop}
\end{eqnarray}
where
\begin{equation}
E_{p} = \sqrt{\vec{p}^{2} + m^{2}},\quad
A(E_{p}) =  \gamma^{0} E_{p} - \vec{\gamma}\cdot
\vec{p} = \gamma^{\mu} A_{\mu} (E_{p}),\quad 
B(E_{p}) = - \gamma^{0} E_{p} - \vec{\gamma}\cdot \vec{p} =
\gamma^{\mu} B_{\mu} (E_{p}).\label{def}
\end{equation}
\begin{center}
\begin{figure}[ht!]
\includegraphics[scale=.8]{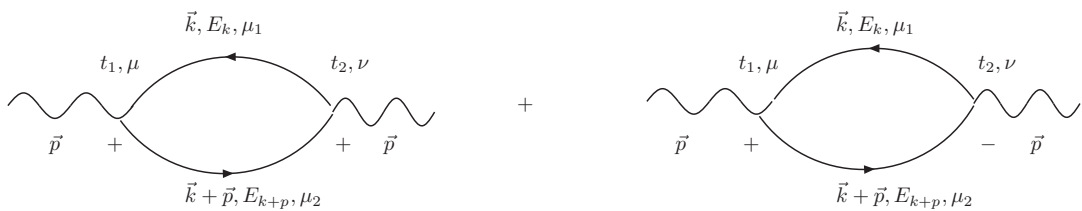}
\caption{The two diagrams contributing to the retarded self-energy for
the photon at one loop.}
\end{figure}
\end{center}
\end{widetext}

The retarded one loop self-energy for the photon (see Fig. 1) can now be calculated
easily. We note here that since the chemical potential occurs as a
phase in the components of the propagator in \eqref{prop}, in the
contribution of the fermion loop to the self-energy at zero
temperature, the dependence on the chemical potential will cancel
out. However, as explained in \cite{silvana3}, for purposes of
applying the thermal operator, we assign distinct chemical potentials
$\mu_{1},\mu_{2}$ to the two fermion propagators in the loop and
identify $\mu_{1}=\mu_{2}=\mu$ only at the end. This simplifies and makes
unambiguous the effect of the thermal operator. In $n$ dimensions in
the mixed space, the
retarded photon self-energy at zero temperature has the form (unfortunately,
both the vector index of the polarization tensor as well as the
chemical potential are conventionally labelled $\mu$, but we do not
believe this will cause any confusion)
\begin{equation}
\Pi_{\rm R}^{\mu\nu (0, \mu)} (t_{1}-t_{2},\vec{p}) = \int \frac{d^{n-1}k}{(2\pi)^{n-1}}\
\varPi_{\rm R}^{\mu\nu (0,\mu)} (t_{1}-t_{2},\vec{p},\vec{k}),\label{integrand}
\end{equation}
where
\begin{widetext}
\begin{equation}
\varPi_{\rm R}^{\mu\nu (0,\mu)}(t_{1}-t_{2},\vec{p},\vec{k}) =
\frac{ine^{2}}{4E_{k}E_{k+p}}\ \theta(t_{1}-t_{2})
e^{-i(\mu_{1}-\mu_{2})(t_{1}-t_{2})} \left[N^{\mu\nu}
  e^{-i(E_{k}+E_{k+p})(t_{1}-t_{2})} - 
  M^{\mu\nu} e^{i(E_{k}+E_{k+p})(t_{1}-t_{2})}\right],\label{mixed}
\end{equation}
with
\begin{eqnarray}
N^{\mu\nu} (E_{k},E_{k+p}) & = & A^{\mu}(E_{k+p})B^{\nu}(E_{k}) -
\eta^{\mu\nu}(A(E_{k+p})\cdot B(E_{k}) - m^{2}) +
A^{\nu}(E_{k+p})B^{\mu}(E_{k}),\nonumber\\
M^{\mu\nu} (E_{k},E_{k+p}) & = & B^{\mu}(E_{k+p})A^{\nu}(E_{k})-\eta^{\mu\nu}
(B(E_{k+p})\cdot A(E_{k})-m^{2}) + B^{\nu}(E_{k+p})A^{\mu}(E_{k}).\label{tensor}
\end{eqnarray}
\end{widetext}

Equation \eqref{mixed} can now be Fourier transformed in the external
time variables to yield ($\omega$, which represents the external
energy, is the variable of Fourier transformation and we
will suppress the arguments
$\vec{p},\vec{k}$ in the self-energy for simplicity)
\begin{widetext}
\begin{eqnarray}
\varPi_{\rm R}^{\mu\nu (0,\mu)} (\omega) & = &
\frac{ne^{2}}{4E_{k}E_{k+p}}\left[-\frac{N^{\mu\nu}}{\omega-E_{k}-\mu_{1}-E_{k+p}+
    \mu_{2}+i\epsilon}
  +
  \frac{M^{\mu\nu}}{\omega+E_{k}-\mu_{1}+E_{k+p}+\mu_{2}+i\epsilon}\right],\label{zeroT}\\
{\rm Im}\ \varPi_{\rm R}^{\mu\nu (0,\mu)} (\omega) & = & \frac{n\pi
  e^{2}}{4E_{k}E_{k+p}}\left[N^{\mu\nu}\delta
  (\omega-E_{k}-\mu_{1}-E_{k+p}+\mu_{2}) - M^{\mu\nu} \delta
  (\omega+E_{k}-\mu_{1}+E_{k+p}+\mu_{2})\right].\label{zeroTimg}
\end{eqnarray}
\end{widetext}
It is clear now that, for a fixed finite value of $\vec{k}$, ${\rm Im}\ 
\varPi_{\rm R}^{\mu\nu (0.\mu)} (\omega)$ vanishes for large values of
$\omega$ and that \eqref{zeroTimg} and \eqref{zeroT} satisfy the zero
temperature dispersion relation \eqref{absorptive}. If we are only interested in the zero 
temperature result, we can set $\mu_{1}=\mu_{2}=\mu$ at this point, which will lead
to the result that the absorptive part of the retarded self-energy and, therefore, the full 
retarded self-energy, at zero temperature do not depend on 
the chemical potential, which is more directly seen from the mixed space result in
\eqref{mixed} (by setting $\mu_{1}=\mu_{2}=\mu$).

As pointed out in \eqref{tor}, at finite temperature, the imaginary
part of the retarded self-enrgy
can be obtained through the application of the thermal operator, which
in the present case takes the form 
\begin{eqnarray}
{\cal O}^{(T,\mu)} & = & \left(1 - \hat{N}_{\rm F}^{(T,\mu_{1})}
  (E_{k}) (1 - S (E_{k}))\right)\nonumber\\
 &\times& \left(1 - \hat{N}_{\rm F}^{(T,\mu_{2})} (E_{k+p}) (1 -
   S(E_{k+p}))\right),\label{TOR}
\end{eqnarray}
where $S(E)$ is a reflection operator that changes $E\rightarrow -E$
and $\hat{N}_{\rm F}^{(T,\mu)} (E)$ denotes a fermion
distribution operator whose action is described in \cite{silvana3}. Applying
the thermal operator \eqref{TOR}, we obtain 
\begin{widetext}
\begin{eqnarray}
{\rm Im}\ \varPi_{\rm R}^{\mu\nu (T,\mu)} (\omega) & = & \frac{n\pi
  e^{2}}{4E_{k}E_{k+p}}\left[\delta (\omega-E_{k}-E_{k+p})\left(1 -
    n_{\rm F}^{+}(E_{k}) - n_{\rm F}^{-}(E_{k+p})\right)
  N^{\mu\nu}\right.\nonumber\\ 
 & - &  \delta(\omega+E_{k}+E_{k+p})\left(1 -n_{\rm
     F}^{-}(E_{k})-n_{\rm F}^{+}(E_{k+p})\right)M^{\mu\nu}
 \label{imgT}\\ 
 & - & \left. \delta(\omega+E_{k}-E_{k+p})\left(n_{\rm
       F}^{-}(E_{k})-n_{\rm F}^{-}(E_{k+p})\right)\bar{N}^{\mu\nu} +
   \delta (\omega-E_{k}+E_{k+p})\left(n_{\rm F}^{+}(E_{k})-n_{\rm
       F}^{+}(E_{k+p})\right)\bar{M}^{\mu\nu}\right],\nonumber 
\end{eqnarray}
where we have used the standard notation $n_{\rm F}^{\pm}(E) = n_{\rm
  F} (E\pm \mu)$ and have defined 
\begin{eqnarray}
\bar{N}^{\mu\nu} & = & N^{\mu\nu} (-E_{k},E_{k+p}) =
A^{\mu}(E_{k+p})A^{\nu}(E_{k}) - \eta^{\mu\nu}(A(E_{k+p})\cdot
A(E_{k})-m^{2}) + A^{\nu}(E_{k+p})A^{\mu}(E_{k}),\nonumber\\ 
\bar{M}^{\mu\nu} & = & M^{\mu\nu} (-E_{k},E_{k+p}) =
B^{\mu}(E_{k+p})B^{\nu}(E_{k}) - \eta^{\mu\nu}(B(E_{k+p})\cdot
B(E_{k}) - m^{2}) + B^{\nu}(E_{k+p})B^{\mu}(E_{k}).\label{tensor1}
\end{eqnarray}
\end{widetext}
 The appearance of new channels of reaction at finite temperature is
 manifest in the absorptive part in \eqref{imgT} and has been obtained
 here from the zero temperature result through the thermal operator
 representation. We note here that while at zero temperature, the
 imaginary part of the retarded photon self-energy leads to the
 probability for the decay of the photon, at finite temperature, the
 additional channels represent the scattering of thermal fermions by a
 photon, which become dominant at very high temperatures (in the hard
 thermal loop approximation).

Using the finite temperature dispersion relation in
\eqref{absorptiveT}, we can now determine the full retarded
self-energy for the photon at finite temperature and chemical
potential from \eqref{imgT} to be 
\begin{widetext}
\begin{eqnarray}
\varPi_{\rm R}^{\mu\nu (T,\mu)} (\omega) & = &
-\frac{ne^{2}}{4E_{k}E_{k+p}}\left[\frac{(1-n_{\rm
      F}^{+}(E_{k})-n_{\rm
      F}^{-}(E_{k+p}))N^{\mu\nu}}{\omega-E_{k}-E_{k+p}+i\epsilon} -
  \frac{(1-n_{\rm F}^{-}(E_{k})-n_{\rm
      F}^{+}(E_{k+p}))M^{\mu\nu}}{\omega+E_{k}+E_{k+p}+i\epsilon}\right.\nonumber\\ 
 & &\qquad \left.-\frac{(n_{\rm F}^{-}(E_{k})-n_{\rm
       F}^{-}(E_{k+p}))\bar{N}^{\mu\nu}}{\omega+E_{k}-E_{k+p}+i\epsilon} + 
\frac{(n_{\rm F}^{+}(E_{k})-n_{\rm
    F}^{+}(E_{k+p}))\bar{M}^{\mu\nu}}{\omega-E_{k}+E_{k+p}+i\epsilon}\right].\label{retardedT}  
 \end{eqnarray}
 \end{widetext}
This demonstrates how starting from only the absorptive part of the
retarded self-energy at zero temperature, we can obtain the full
retarded self-energy at finite temperature and chemical potential
through the use of the dispersion relation and the application of the
thermal operator. For $n=4$, Eq. \eqref{retardedT} reduces to the
well known result in  QED \cite{silvana3,kapusta,lebellac}. We note
here that both \eqref{imgT} as well as \eqref{retardedT} are
non-analytic at the origin in the energy-momentum space because of the
additional channels of reaction. The
non-commuting nature of the limits $\omega\rightarrow 0,
\vec{p}\rightarrow 0$ and $\vec{p}\rightarrow 0,\omega\rightarrow 0$
arises because they represent different physical effects at finite
temperature. However, for
$\vec{p}\neq 0$, the retarded self-energy $\Pi_{\rm R}^{\mu\nu
(T,\mu)} (\omega)$ is an analytic function in the entire upper half of
the complex $\omega$-plane which justifies the dispersion relation in
\eqref{absorptiveT}. 

Let us next consider the Schwinger
model \cite{schwinger1} which corresponds to two dimensional massless QED. For $m=0$, in
two dimensions ($n=2$) we have various simplifications. First, we can write
\begin{equation}
E_{k} = |k^{1}|,\quad E_{k+p} = |k^{1}+p^{1}|.\label{energy}
\end{equation}
Furthermore, in two dimensions the tensors in \eqref{tensor} and
\eqref{tensor1} simplify to have the forms
\begin{widetext}
\begin{eqnarray}
N^{\mu\nu} & = &
-2E_{k}E_{k+p}\left[\theta(k^{1})\theta(-k^{1}-p^{1})u^{\mu}_{+}u^{\nu}_{+} +
  \theta (-k^{1})\theta(k^{1}+p^{1})u^{\mu}_{-}u^{\nu}_{-}\right],\nonumber\\
M^{\mu\nu} & = &
-2E_{k}E_{k+p}\left[\theta(-k^{1})\theta(k^{1}+p^{1})u^{\mu}_{+}u^{\nu}_{+} +
  \theta(k^{1})\theta(-k^{1}-p^{1})u^{\mu}_{-}u^{\nu}_{-}\right],\nonumber\\
\bar{N}^{\mu\nu} & = &
2E_{k}E_{k+p}\left[\theta(-k^{1})\theta(-k^{1}-p^{1})u^{\mu}_{+}u^{\nu}_{+} +
  \theta(k^{1})\theta(k^{1}+p^{1})u^{\mu}_{-}u^{\nu}_{-}\right],\nonumber\\
\bar{M}^{\mu\nu} & = &
2E_{k}E_{k+p}\left[\theta(k^{1})\theta(k^{1}+p^{1})u^{\mu}_{+}u^{\nu}_{+} +
  \theta(-k^{1})\theta(-k^{1}-p^{1})u^{\mu}_{-}u^{\nu}_{-}\right],\label{identity}
\end{eqnarray}
\end{widetext}
where we have defined the null vectors
\begin{equation}
u^{\mu}_{+} = (1, -1),\quad u^{\mu}_{-} = (1,1).\label{null}
\end{equation}

With the relations \eqref{energy} and \eqref{identity}, the
temperature dependent part of ${\rm Im}\ \varPi_{\rm R}^{\mu\nu
  (T,\mu)}(\omega)$ in \eqref{imgT} can be simplified and takes the
form (we use the 
standard notation \cite{dasbook,adilson} $A^{(T)} = A^{(0)} + A^{(\beta)}$ decomposing any
observable to its zero temperature part and the temperature dependent
part)
\begin{widetext}
\begin{eqnarray}
\lefteqn{{\rm Im}\ \varPi_{\rm R}^{\mu\nu (\beta,\mu)}(\omega)}\nonumber\\
 & = & \pi
e^{2}\left[\delta(\omega+p^{1})u^{\mu}_{+}u^{\nu}_{+}\left\{\theta(k^{1})n_{\rm
      F}^{+}(E_{k}) - \theta(-k^{1})n_{\rm
      F}^{-}(E_{k})-\theta(k^{1}+p^{1})n_{\rm F}^{+}(E_{k+p}) +
    \theta(-k^{1}-p^{1})n_{\rm
      F}^{-}(E_{k+p})\right\}\right.\nonumber\\
&+&\left.\delta(\omega-p^{1})u_{-}^{\mu}u_{-}^{\nu}\left\{\theta(-k^{1})n_{\rm
      F}^{+}(E_{k}) - \theta(k^{1})n_{\rm
      F}^{-}(E_{k})-\theta(-k^{1}-p^{1})n_{\rm
      F}^{+}(E_{k+p})+\theta(k^{1}+p^{1})n_{\rm
      F}^{-}(E_{k+p})\right\}\right].\label{schwinger}
\end{eqnarray}
\end{widetext}
If we use the fact that $\varPi_{\rm R}^{\mu\nu}$ is the integrand in an
integral involving $k^{1}$ for the self-energy (see, for example,
\eqref{integrand}), we can redefine $k^{1}\rightarrow -k^{1}-p^{1}$ in
some of the terms in \eqref{schwinger} to rewrite the temperature
dependent part as
\begin{eqnarray}
{\rm Im}\ \varPi_{\rm R}^{\mu\nu (\beta,\mu)}(\omega) & = & \pi e^{2}
\epsilon(k^{1}) \left(n_{\rm F}^{+}(E_{k})+n_{\rm
    F}^{-}(E_{k})\right)\label{imgschwinger}\\
&\times&\!\!\!\left(\delta(\omega+p^{1}) u^{\mu}_{+}u^{\nu}_{+} -
  \delta(\omega-p^{1})u^{\mu}_{-}u^{\nu}_{-}\right),\nonumber
\end{eqnarray}
where $\epsilon(k^{1})=\theta(k^{1})-\theta(-k^{1})$. The important
thing to note here is that
the integrand of the 
imaginary part of the temperature dependent retarded
self-energy is anti-symmetric in the integration variable $k^{1}$
because of the alternating step function. As a result, through the
dispersion relation \eqref{absorptiveT}, the temperature dependent
part of the complete retarded self-energy, $\varPi_{\rm R}^{\mu\nu
  (\beta,\mu)}(\omega)$, would also
inherit this anti-symmetry. It follows, therefore, that the
temperature dependent imaginary part of the retarded self-energy as
well as the retarded self-energy vanish (when integrated over $k^{1}$) for
the Schwinger model. This result is a generalization of \cite{adilson}
to the case of a nonzero chemical potential. We note here that the delta
function structure as well as the manifest 
anti-symmetry in \eqref{imgschwinger} is a reflection of helicity
conservation for massless fermions scattering from a photon background
in $1+1$ dimensions.

\vskip 1cm

\noindent{\bf Acknowledgment:}
\medskip

One of us (AD) acknowledges the Fulbright Foundation for a
fellowship. This work was
supported in part by US DOE Grant number DE-FG 02-91ER40685, by CNPq
and by FAPESP, Brazil. We have used the program Jaxodraw \cite{binosi}
for generating the figure in this paper.

\end{document}